\PassOptionsToPackage{hyphens}{url}
\documentclass{sigchi}

\toappear{\scriptsize Permission to make digital or hard copies of all or part of this work for personal or classroom use is granted without fee provided that copies are not made or distributed for profit or commercial advantage and that copies bear this notice and the full citation on the first page. Copyrights for components of this work owned by others than the author(s) must be honored. Abstracting with credit is permitted. To copy otherwise, or republish, to post on servers or to redistribute to lists, requires prior specific permission and/or a fee. Request permissions from permissions@acm.org. \\ {\emph{CHI '20, April 25--30, 2020, Honolulu, HI, USA.} } \\
\textcopyright~2020 Copyright is held by the owner/author(s). Publication rights licensed to ACM. \\ ACM ISBN 978-1-4503-6708-0/20/04\ ...\$15.00.\\
http://dx.doi.org/10.1145/3313831.3376446 }

\usepackage{balance}       %
\usepackage{graphics}      %
\usepackage[T1]{fontenc}   %
\usepackage{txfonts}
\usepackage{mathptmx}
\usepackage[pdflang={en-US},pdftex]{hyperref}
\usepackage{color}
\usepackage{booktabs}
\usepackage{textcomp}

\usepackage{xspace}
\usepackage{multirow}
\usepackage{float} %
\usepackage[separate-uncertainty = true]{siunitx}
\usepackage[subtle]{savetrees}
\usepackage{tabularx}
\usepackage{dcolumn}
\usepackage{nameref}
\usepackage{multirow}

\newcolumntype{d}[1]{D{.}{.}{#1}}

\newcolumntype{b}{X}
\newcolumntype{m}{>{\hsize=.39\hsize}X}
\newcolumntype{h}{>{\hsize=.7\hsize}X}

\usepackage{microtype}        %
\usepackage{ccicons}          %

\newcommand{\labyrinth}{\emph{Labyrinth}\xspace}
\newcommand{\nsamples}{46\xspace}
\newcommand{\nfeatures}{75\xspace}
\newcommand{\pusher}{\emph{pusher}\xspace}
\newcommand{\collector}{\emph{collector}\xspace}

\newcommand{\svm}{\emph{SVM}\xspace}
\newcommand{\rf}{\emph{RF}\xspace}
\newcommand{\cv}{\emph{CV}\xspace}

\newcommand{\rtwo}{\emph{R$^2$}\xspace}

\newcommand{\fone}{\emph{F$_1$}\xspace}

\newcommand{\chronemics}{\emph{chronemics}\xspace}
\newcommand{\commcontent}{\emph{communication content}\xspace}

\newcommand{\events}{\emph{events}\xspace}

\newcommand{\selfreport}{\emph{self-report}\xspace}
\newcommand{\bestfeatures}{\emph{bestfeatures}\xspace}

\newcommand{\bfval}[1]{$BF_{10}=#1$}
\newcommand{\cohensd}[1]{\emph{d}~=~$#1$}

\newcommand{\rqone}{\emph{RQ1}\xspace}
\newcommand{\rqtwo}{\emph{RQ2}\xspace}
\newcommand{\rqthree}{\emph{RQ3}\xspace}

\everypar{\looseness=-1}

\def\plaintitle{Recognizing Affiliation: Using Behavioural Traces to Predict the Quality of Social Interactions in Online Games}
\def\plainauthor{Julian Frommel, Valentin Sagl, Ansgar E.\ Depping, Colby Johanson, Matthew K.\ Miller, Regan L.\ Mandryk}

\def\plainkeywords{affiliation; social interaction; evaluation; prediction; recognition; cooperative games; machine learning; bonding}

\makeatletter
\def\url@leostyle{%
  \@ifundefined{selectfont}{
    \def\UrlFont{\sf}
  }{
    \def\UrlFont{\small\bf\ttfamily}
  }}
\makeatother
\def\pprw{8.5in}
\def\pprh{11in}

\definecolor{linkColor}{RGB}{6,125,233}
\hypersetup{%
  pdftitle={\plaintitle},
  pdfauthor={\plainauthor},
  pdfkeywords={\plainkeywords},
  pdfdisplaydoctitle=true, %
  bookmarksnumbered,
  pdfstartview={FitH},
  colorlinks,
  citecolor=black,
  filecolor=black,
  linkcolor=black,
  urlcolor=linkColor,
  breaklinks=true,
  hypertexnames=false
}

\begin{document}

\title{\plaintitle}

\numberofauthors{1}
\author{%
    \alignauthor{Julian Frommel\textsuperscript{1,2}, Valentin Sagl\textsuperscript{2}, Ansgar E.\ Depping\textsuperscript{2}, Colby Johanson\textsuperscript{2},\\ Matthew K.\ Miller\textsuperscript{2}, Regan L.\ Mandryk\textsuperscript{2}\\
    \affaddr{\textsuperscript{1}Institute of Media Informatics, Ulm University, Ulm, Germany}\\
    \affaddr{\textsuperscript{2}Department of Computer Science, University of Saskatchewan, Saskatoon, SK, Canada}\\
    \email{\{firstname.lastname\}@usask.ca}}\\
}

\maketitle

\begin{abstract}
Online social interactions in multiplayer games can be supportive and positive or toxic and harmful; however, few methods can easily assess interpersonal interaction quality in games. We use behavioural traces to predict affiliation between dyadic strangers, facilitated through their social interactions in an online gaming setting. 
We collected audio, video, in-game, and self-report data from 23 dyads, extracted 75 features, trained Random Forest and Support Vector Machine models, and evaluated their performance predicting binary (high/low) as well as continuous affiliation toward a partner. The models can predict both binary and continuous affiliation with up to 79.1\% accuracy ($F_1$) and 20.1\% explained variance ($R^2$) on unseen data, with features based on verbal communication demonstrating the highest potential. Our findings can inform the design of multiplayer games and game communities, and guide the development of systems for matchmaking and mitigating toxic behaviour in online games.
\end{abstract}


{}


\section{Introduction}

Multiplayer games are popular among players, with recent statistics showing that frequent gamers play with others online for 6 hours a week on average \cite{esa2017}. Social aspects of gaming are among the main motivators to play online multiplayer games \cite{de2017validating, frostling2009first, jansz2005gaming, jansz2007appeal}, which leads to a lot of social interactions between players. Social interactions in online games have been linked with many positive outcomes, including the emergence of social capital \cite{huvila2010social, steinkuehler2006everybody}, forming and extending physical world relationships \cite{SCHIANO201465, trepte2012social, williams2006tree}, and positive effects on psychological well-being \cite{depping2018designing}; however, harmful or toxic interactions can also occur, which have negative ramifications for player experience \cite{Foo:2004:DGP:1067343.1067375, kwak2015exploring} and the overall health of game communities. Game developers spend time and effort addressing toxic game environments and  communities \cite{kaplan2017developer, pcgamerPrescottOverwatch}; examples such as tribunals in \emph{League of Legends} \cite{lolKotakuTribunal} highlight the difficulties of assessing the quality of social interactions that occur between players and ensuring positive game communities.

To create and include game mechanics and community features that promote positive social interactions between players, developers must first be able to \emph{evaluate the quality of social interactions} in their game; however, methods to do so are limited. %
Self-report measures assessing players' subjective experience of social interactions exist \cite{johnson1982measurement, larzelere1980dyadic, rempel1985trust}, but are non-automated and hindered by guessing behaviour \cite{Westera2016}, retrospective bias \cite{mauss2009measures}, social desirability \cite{mauss2009measures}, and disruptiveness when administered during gameplay \cite{frommel2015integrated}. Generally, questionnaires and interviews are impractical to administer after a game's release. Using behavioural traces to predict self-reported experiences allows continuous, real-time, and unobtrusive assessment~\cite{frommel2018towards}.
Many such methods are directed at important experiences in single-player settings (e.g., frustration, boredom, fun)~\cite{cowie2001emotion, fragopanagos2005emotion, MANDRYK2007329, Mandryk:2006:COE:1124772.1124926}, but limited for evaluating the social aspects of play that are important in multiplayer settings.
As such, there is a need for researchers to develop methods for evaluating the quality of social interactions in multiplayer games in a way that is practically applicable---that is, using behaviour (in contrast to disruptive self-report measures) and unobtrusive sensors that do not interfere with the game experience or the social interactions between players.
In this paper, we build computational models of the quality of social interactions, by mining players' behavioural traces for cooperative dyads as a first step of such an assessment in multiplayer games.
A variety of criteria are relevant for evaluating how players experience social interactions, such as the perception of cooperation and interdependence~\cite{depping2017cooperation, Harris:2019:ABI:3290605.3300239}, trust~\cite{depping2017cooperation, Depping:2016:TMS:2967934.2968097}, and supportiveness~\cite{Kou:2014:PSU:2658537.2658538}. We assess these qualities using the umbrella term \emph{affiliation} and measure it with an 11-item scale used in previous work on social closeness~\cite{depping2017cooperation}. By predicting questionnaire responses with multiple criteria, we use player behaviour to differentiate between good and bad interactions. In addition to our main multimodal approach, prediction using single feature categories such as game performance could be beneficial in cases when data sources are limited, e.g., as players do not share their video because of privacy. 
As such, this paper aims to answer the following research questions:

\begin{itemize}
    \item \rqone: Is it possible to predict a player's affiliation toward a co-player in an online multiplayer game setting from unobtrusively gathered behavioural data?
    \item \rqtwo: How do models using only features from a single category (e.g., game performance, facial expressions) perform?
    \item \rqthree: Which behavioural traces are important features for predicting affiliation during play? 
\end{itemize}

We conducted an online study with 46 participants, in which strangers were matched up in pairs and played a cooperative digital game together. During the game, participants were connected via audio chat for communication. Audio recordings, video recordings of the participants' webcams while playing, 
in-game log data, and self-reported data were collected. As we are interested in a method that is practical for use in real gameplay settings, we focused on data that can be collected unobtrusively during gameplay, and avoided categories of features that required specialized hardware (e.g., physiological sensors based on skin contact) or explicit player input required in the moment of play (e.g., state-based self report measures). 
After playing, participants reported their affiliation toward each other. We employed a supervised machine learning approach, training models with the aim of predicting the participants' self-reported level of affiliation. 

Our results demonstrate that predicting affiliation using behavioural traces is possible with up to 79.1\% accuracy ($F_1$) for binary classification and 20.1\% explained variance ($R^2$) for continuous measures. Further, an analysis of models using category-based feature subsets (e.g., in-game performance) shows that models based on verbal communication features (\chronemics and \commcontent) perform best, demonstrating their high value for prediction. Finally, an analysis of feature importances gives first insights into the connection between player behaviour and social interaction quality, which can inform future hypotheses for controlled experiments studying causal relationships. %

These findings can help researchers and practitioners who want to evaluate social interaction quality in online multiplayer games. %
Applications include game evaluation, mitigating toxic behaviour in published games, and improving matchmaking.
This research is critical because the gaming industry increasingly trends toward games as a service (cf.~\cite{cook2014gaas}), in which publisher revenues and player experiences both depend on healthy, ongoing communities.
Our approach can help developers who require community health monitoring tools to identify shifts in their communities, evaluate new features, and tweak and optimize existing features.

\section{Background}
\label{sec:background}
Our work is related to assessment approaches in gaming and to research on the relationship between social interactions and human behaviour, which informs the behavioural traces that we use as features.

\subsection{Using Behaviour to Assess Social Interaction Quality}
In this paper, we evaluate the feasibility of measuring affiliation in dyads, where we consider interpersonal trust as important~\cite{depping2017cooperation}. 
Trust has been used to characterize social relationships in computer-mediated communication~\cite{jarvenpaa1999communication, johnson1982measurement, larzelere1980dyadic, rempel1985trust, rusman2010fostering}, game settings~~\cite{Depping:2016:TMS:2967934.2968097}, and social closeness in multiplayer games specifically~\cite{depping2017cooperation}.
While there are questionnaires that measure self-reported trust \cite{johnson1982measurement, larzelere1980dyadic, rempel1985trust}, there is little previous work on  %
automated and unobtrusive methods affiliation assessment.

The detection of affiliation is, however, closely related to \emph{emotion recognition} \cite{cowie2001emotion,fragopanagos2005emotion}, as it can be considered a method of measuring a user's psychological state based on implicit signals. In a gaming context, the detection of emotion can be used to evaluate the quality of experiences \cite{mandryk2005modeling} or to adapt game features based on players' emotional states \cite{chanel2008boredom, chanel2011emotion, liu2009dynamic, negini2014using, schrader2017rising, tijs2008creating, tijs2008dynamic}. Previous work has shown a relationship between emotion and trust~\cite{williams2001whom}, especially toward unfamiliar people \cite{dunn2005feeling}, suggesting value of features used in emotion recognition for the evaluation of affiliation in multiplayer games. %
While emotion recognition methods can inform our feature selection, in general they are rarely used to evaluate multiplayer settings (e.g., \cite{mandryk2004physiological}), hinting at a lack of guidance on the assessment of social experiences. 

On the other hand, the analysis of text messages to detect toxic behaviour in games \cite{martens2015toxicity, THOMPSON2017149} is related to the negative consequences of harmful in-game social interactions. While these methods can be used to evaluate social interactions occurring in multiplayer settings and thus inform our use of content-based conversational features, they are not helpful for assessing positive outcomes of beneficial social interactions and they further rely on objective criteria on what is considered toxic. However, different players can experience interactions or messages very differently.

In summary, we find that there is a lack of guidance on methods for the assessment of social interaction quality. Approaches to modeling affiliation between players should consider how a player experiences an interaction, not just the observable characteristics of the interaction itself. Therefore, in this paper, we examine whether players' behaviour can be used to detect affiliation as experienced by players.

\subsection{Potential Indicators of Affiliation}
We rely on previous work studying the relationship between human behaviour and  affiliation to inform which behavioural traces might be useful for the assessment of social closeness. Affiliation is important for understanding player behaviour~\cite{Poeller:2018:LMI:3173574.3173764}, as it is a central motive for human behaviour according to Motive Disposition Theory~\cite{mcclelland1985motives}. However, we aim to predict how players perceive social interactions comprised of a broad spectrum of qualities that we have summarized under the umbrella term \textit{affiliation}. To inform potentially relevant features, we build on literature from other fields, such as work and organizational psychology. In particular, we are interested in traces that %
can be collected unobtrusively with low-fidelity sensors in a natural gaming setting. We use features that are related to affiliation and features that can be used for emotion recognition due to the established relationship between emotion and social closeness~\cite{dunn2005feeling, williams2001whom}. %

Depping and Mandryk found that games requiring interdependence between players can build trust, an effect that was fully explained by a greater number of conversational turns during interdependent play \cite{depping2017cooperation}. This suggests that features related to the timing of conversation should be explored as a predictor of social interaction quality among players. Chronemic conversational analysis (i.e., analyzing the timing of conversations) in a game context may require different considerations than in productive or serious contexts. For example, past work has shown that the action of playing a game itself can be considered a conversational act \cite{429-chess-as-a-conversation}; a gap in game-related conversation may therefore reflect increased communication through in-game moves, and a speaker who receives no verbal response may see the answer reflected in the other player's in-game actions. Occurrences that might be considered negative in serious or productive contexts may be a normal part of communication between players in a game. %

A major part of a social interaction is the communication between humans. In earlier work, Gilbert et al.~\cite{gilbert2009predicting} used the content of communication in models to predict social ties between social network users. 
As we expect a similar connection between communication content and the quality of the social interactions in a game context, we use it as a feature for our model.
Previous research showed that eye blink rate (EBR) is a non-invasive indicator for central dopamine function \cite{jongkees2016spontaneous}. Research suggests a link of dopamine to emotions~\cite{badgaiyan2010dopamine, nieoullon2003dopamine} and reward function through social interactions~\cite{vrticka2012interpersonal, vrticka_neuroscience_2012}. %
Therefore, we consider EBR a relevant feature for our model.
Due to the importance of emotions, we include facial expressions features based on previous work showing that facial action units \cite{ekmanfacial} can be used to detect emotional state \cite{bartlett1999face, mcduff2016affdex, senechal2013smile}. %

Many games use challenges to elicit fun \cite{lopes2011adaptivity, malone1980makes}, suggesting that player performance and experience are connected. Players' in-game performance was used in previous research on gameplay adaptation \cite{hunicke2005case} and emotion recognition in games \cite{frommel2018towards}, hinting at a potential importance for affiliation between players. We suspect that the utility of in-game measures might be high in multiplayer settings, in which players' performance does not depend solely on themselves. In addition, we consider players' in-game actions relevant as they are potentially linked to their experience \cite{yannakakis2011experience}. As such, they might be informative for the assessment of affiliation and are therefore used in our model.

Besides features derived from players' behaviour, we use a small set of relatively stable traits that can be collected outside of a game session through one-time self-reports. %
In particular, we added features based on previous work suggesting that perception of the other, e.g., trust, is affected by age \cite{CLARK2013361}, gender \cite{croson2009gender}, the gender combination of people involved in an interaction \cite{bevelander2011ms, SCHARLEMANN2001617, sun2007effect, sun2007gender}, and personality traits like agreeableness or propensity to trust \cite{evans2008survey, mooradian2006trusts}. 
Finally, we use features based on identification as a gamer and preferred gaming style because of previous work suggesting a link between gaming frequency and social interaction \cite{williams2006groups}. 
\vspace{-3pt}
\section{Data Collection}

We collected behavioural and self-report data in an online study of pairs of strangers playing a networked, cooperative, and interdependent game to create predictive models of affiliation.
\vspace{-3pt}
\subsection{Game: Labyrinth}
\begin{figure}
    \centering
	\includegraphics[width=.75\columnwidth]{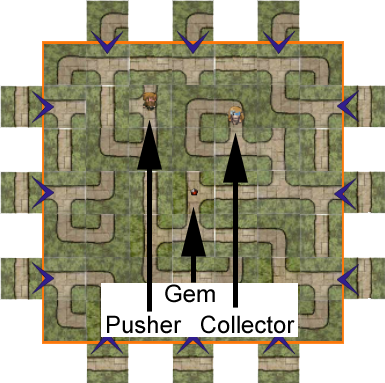}
	\caption{Labyrinth is a collaborative online game in which two players have to work together to collect gems using complementary roles.}
	\label{fig:labyrinth}
\end{figure}

We used \labyrinth (see Figure~\ref{fig:labyrinth}), a digital online multiplayer game based on a similarly named board game \cite{labyrinth}, which has been used in previous research studying social closeness in games \cite{depping2017cooperation, Depping:2016:TMS:2967934.2968097}. While the game has multiple modes, we used the cooperative and interdependent version of the game as both of these mechanics are important for social closeness between players \cite{depping2017cooperation} and we assume that a minimum required level of affiliation must be built between players to build a computational model. In the game, players have to cooperate to gather collectibles (gems) by rearranging a maze, operating under a fixed time limit. The players have different roles: only the \collector can pick up the gems to increase the shared score and only the \pusher can push walls to rearrange the maze and create paths for the \collector. These game mechanics lead to cooperative and interdependent gameplay through shared goals and tightly-coupled complementary roles. Players communicate via audio chat during the game.

\subsection{System}
\labyrinth was developed in Unity \cite{unity} and presented online using WebGL. Players were connected to a game server that managed synchronized game logic using the Unity Multiplayer HLAPI \cite{unityHLAPI} and WebSockets. The audio chat used WebRTC, a peer-to-peer communication standard for web browsers. In our setup, a server running Kurento Media Server \cite{Lopez:2016:KWM:2964284.2973798} was used instead of a peer-to-peer connection. This increased reliability when connecting users and allowed recording the participants' audio streams.

\subsection{Study Setting}
The study was conducted in an online setting that balanced the need for a natural gaming environment with the need for experimental control. By conducting the study online, we were able to evaluate the social interactions between two strangers, both playing at home, who were matched by a matchmaking system---a natural scenario that happens in many commercial online games. Although we considered gathering data in the laboratory, the idea that local participants might know of each other, move in the same social circles, or interact under the assumption that they may meet again %
led to our decision to conduct the study online, avoiding uncontrollable effects of these factors on their interactions. %

\subsection{Procedure}
The study was conducted using Amazon's Mechanical Turk (MTurk), a human-intelligence task market that has been shown to provide reliable data for HCI studies when special care is taken to verify data quality \cite{buhrmester2011amt, kittur2008crowdsourcing, mason2012conducting, paolacci2010running}. Participants connected to a web server hosting the study. They were instructed on the procedure of the study and were required to provide informed consent. They completed an initial questionnaire measuring demographics, propensity to trust, and personality. Next, they watched a tutorial video explaining the study and how to play \labyrinth. They were then prompted to allow webcam and microphone access. A set of guidelines was shown reminding participants to ensure a good video (e.g., look at the screen), along with a live preview of their video to provide feedback. Note that participants did not see each others' videos---we gathered them for our own data analyses, but video previews were not displayed during gameplay to their partner or to themselves, due to potential effects on social interaction \cite{Miller:2017:TLG:3025453.3025548, tscholl2005effect} and as neither is common in multiplayer games. They were then matched randomly with another participant, connected to each other via audio chat, and redirected to the \labyrinth game page. 
The game roles were assigned and players were instructed via text on their role and how to start the game, i.e., indicating that they were ready to start, after both players had loaded the game.
They then played \labyrinth for five minutes, received feedback on their performance, changed roles, and played a second round for an additional five minutes. Subsequently, they were disconnected and completed a concluding questionnaire about their experience.
\subsection{Measures}
Before the game, we measured the participants' propensity to trust as a trait using the General Trust Scale \cite{Yamagishi1994}, their Big 5 personality traits using the Ten-Item Personality Inventory \cite{ehrhart2009testing}, and their age and gender. In addition, they reported their self-identification as a gamer on a 100-point scale \cite{mandryk2017toward}, enjoyment of game genres using a checklist, and BrainHex player type \cite{nacke2014brainhex} by choosing their preferred play style. %
Behavioural traces were based on data recorded during gameplay. We recorded the communication of participants through the audio channel on the server as synchronized raw audio files from start to end of the audio connection for both participants. Additionally, the participants' webcam video, in-game events, and in-game performance (i.e., scored points) were recorded. %
After the game, we measured players' affiliation using an 11-item scale based on items from other scales \cite{johnson1982measurement, larzelere1980dyadic, rempel1985trust}. This scale was used to measure trust in games \cite{depping2017cooperation, Depping:2016:TMS:2967934.2968097}, but covers other aspects such as honesty, fairness, and reliability, which we summarize as \emph{affiliation}. 
We tested the scale on our data and found it to have excellent internal consistency ($\alpha$~=~.952) \cite{george1999spss}.

\subsection{Participants}
Participants were recruited using MTurk. %
Due to the nature of the task (matching and recording participants), it was easy to detect bots or participants who did not diligently complete the task, allowing for a simple assessment of data validity. All data (audio, video, game, questionnaire responses) were manually inspected and pairs of participants were removed if the inspection suggested that participants did not actually complete the full study or play with each other as intended. In particular, both of them had to have completed all questionnaires and scored at least one point, indicating attempted gameplay (as scoring a point was only possible if both participants played). They had to be connected via audio chat, which was verified by the presence of both audio files, which were created after the audio connection was closed and as such were only present when participants were successfully connected. In the end, we had valid data for 46 participants (female~=~16, male~=~30) aged 21 to 58 (\emph{M}~=~34.35, \emph{SD}~=~9.45). There were 23 pairs in the study (female--female~=~2, male--male~=~9, female--male~=~12). %

\section{Prediction Models}
We used the study data to generate prediction models.

\subsection{Approach}

As there has been little previous work guiding the implementation of a behaviour-based assessment of affiliation between players, we follow an exploratory approach. We collected features that we considered potentially related to players' perception of the interaction, trained machine learning models, and evaluate their performance in predicting self-reported affiliation---a supervised learning task similar to earlier work predicting human ratings~\cite{frommel2018towards, Summerville:2017:UME:3102071.3102080}. 
Depending on the form of the outcome variable, i.e., what is being predicted, supervised learning tasks are tackled with classification (prediction of classes) and regression (prediction of continuous values) approaches. %

In this paper, we operationalize the quality of social interactions through a self-reported measure of affiliation; we calculate a single score from the 11-item scale following previous work \cite{depping2017cooperation, Depping:2016:TMS:2967934.2968097}, yielding a continuous affiliation score (range: 1--7, $Mdn=5.46$ $M=5.31$, $SD=1.28$), which makes prediction on this scale a regression task. Predicting binary classes (i.e., if a player experiences low or high affiliation) is potentially easier than predicting exact scores, but can still suffice in practical applications. Thus, we also evaluate binary classification, with the goal of predicting \emph{low} or \emph{high} affiliation. However, there is no general objective criterion dividing the scale into population norms for low and high states of affiliation. As such, to evaluate binary classification, we used a median split on our distribution to generate low and high categories of affiliation and evaluate binary predictions employing a binary classifications approach (\emph{N$_{low}$}~=~24,  \emph{N$_{high}$}~=~22).%

As we collected self-reported affiliation once, we only have a single label for each gameplay phase. Thus, we consider the data of a single player as a single sample, and calculate a variety of higher-order features describing the whole gameplay phase. A sample then consists of a feature vector for the whole gameplay phase and a corresponding affiliation label. Generally, we have a unique sample for each participant, but use a small set of features from data of both participants (e.g., communication). 
We take this approach because we assume that players' experience of an interaction can diverge and we want to predict affiliation for each individual player. %
We tackle the potential correlation of affiliation scores within dyads in our cross-validation approach (described later). %
This approach allows us to use features from various sources as a combined input. %
While explicit time information of features is discarded, time information can still be used in features, e.g., for time of silence. %
Even though this approach is not necessarily optimal for separate data streams, it allows us to combine the different types of data streams into a single model, which makes it easy to evaluate the performance of different combinations of features, making it appropriate for addressing our research questions. %

\vspace{-4pt}
\subsection{Data Preprocessing}
Because our study was conducted online in participants' homes, some aspects of the collected data were missing or unusable for some participants. As previously reported, we removed all pairs who did not actually play the game, whose questionnaire responses were invalid, or who did not successfully establish an audio connection. However, this does not imply a minimum amount of communication between the players (in the extreme case a participant could mute their audio despite the instruction to enable sound). We inspected the participants' audio and video files to identify issues that would make them unusable, for example, poor framing or lighting, or the players not talking. Audio files were filtered to remove background noise, such as other people talking in the background, and synchronization was verified and adjusted if necessary. The framerates of videos were tested and standardized. 

\subsection{Feature Generation}
Although all participants for which we generated features were considered valid, single features could be invalid due to a single unusable data stream (e.g., video was unusable or the players did not talk). We retained the sample, marked single features as missing, and imputed them based on existing data at a later step to avoid losing too much valid data. 
In general, we generated \nfeatures features that can be associated with seven categories (see Table~\ref{tab:featureList}). All features were generated in multiple steps: the source data was preprocessed and then basic features were extracted. Features were selected based on previous work hinting at an importance for social interaction quality (see Section~\nameref{sec:background}). %

\textbf{Chronemics:}
Previous work has found that the number of conversational turns predicts trust formation in games \cite{depping2017cooperation}; however, conversational turns alone are a coarse reflection of the balance and timing in a conversation. To capture more detail about the timing of the conversation, we included a total of 12 chronemic features that were processed using custom software written in Python \cite{Rossum:1995:PRM:869369}.
Using the cleaned audio files, we started by splitting each participant's audio into segments of speaking and pausing using amplitude-based thresholding. This yielded measures of total time speaking, count of speech segments, average length of a speech segment, average length of pause segment, and \emph{SD} of length of speech segment.
Next, we analyzed each pair's audio files together and calculated conversational turns by merging pausing and speaking segments that were uninterrupted by a speaking segment from the other participant. We also analyzed silence in the conversation by comparing the pausing segments of the two participants. We calculated the total amount of silence in the conversation, the average length of each period of silence, the fraction of the conversation spent in silence, and the length of first silence of the call (before either person spoke).
Finally, we included two boolean features for each participant indicating whether they were the dominant speaker in the conversation, one generated from conversational turns and the second from speech time.

\textbf{Communication Content:}
We generated 17 features relating to the content of players' communication on a semantic level, based on word counts. Audio files were transcribed and then semantically analyzed using the Linguistic Inquiry Word Count (LIWC) tool \cite{pennebaker2015development}. We used summary dimensions (\emph{Total Word Count}, \emph{Analytic}, \emph{Clout}, \emph{Authentic}, \emph{Tone}), personal pronouns that could indicate players seeing themselves as single players or as a team (\emph{I}, \emph{You}, \emph{We}), general dimensions related to social closeness (\emph{Social}, \emph{Affiliation}), dimensions that could be related to gameplay and scoring (\emph{Motion}, \emph{Space}, \emph{Time}, \emph{Number}), and affect dimensions (\emph{Affect}, \emph{Positive Emotions}, \emph{Negative Emotions}). 

\textbf{Eye Blink:}
We employed a custom video-based blink detection tool using Python \cite{Rossum:1995:PRM:869369}, OpenCV \cite{opencv_library}, and Keras \cite{chollet2015keras} to count blinks. Frames of the videos were analyzed by extracting images of the eyes and using a convolutional neural network predicting if the eyes are opened or closed. This network was trained on the ``The Closed Eyes in the Wild'' database \cite{eyesInTheWildDB}, consisting of data from 2423 subjects with labelled images of faces and opened and closed eyes. We validated the recognition using 3 short test videos and manually labelled frames. The algorithm achieved an $F_1$ score of .989 on those videos, suggesting good performance even if our participants' videos were noisier than test data. Participant videos were manually inspected and videos that were problematic for the blink detection, e.g., participants with glasses, were marked invalid. %
As a result, the EBR features of 12 people were discarded. For the remaining participants, blinks per second were calculated; to reduce noise, a moving average window (five seconds) was used. Blinks per minute were also calculated and the average EBR (as well as its \emph{SD}) was calculated over the duration of the video. Data were marked invalid if the face could not be detected in more than 10\% of frames or there was no phase of continuous detection over 5 minutes, which is considered a suitable time span for EBR in experimental designs \cite{jongkees2016spontaneous}.%

\textbf{Facial Expressions:}
We added 16 features related to the players' emotions, based on their facial expressions. We used the \emph{AFFDEX SDK} \cite{mcduff2016affdex} to predict emotional state based on facial expressions. The SDK generates confidence scores between 0 and 100 in each frame for engagement, contempt, surprise, anger, sadness, disgust, fear, and joy, representing the strength of each emotion reflected in the players' face for that frame. For each emotion, we calculated two features over the whole gameplay duration: 
a general measure of overall strength using the average predicted strength over all frames, and a count of strength peaks, defined as local maxima over a threshold of 50, to better reflect facial expressions of a short duration (cf. \cite{ekman1992argument}). 
As the SDK only generates predictions when the face is visible, we calculated the ratio of frames for which the SDK recognized a face compared to the overall number of frames and considered all facial expression features for a sample only valid if this ratio was over 80\%.   

\textbf{In-Game Performance:}
We calculated 12 features relating to in-game performance based on score, as we expected that a player's impression of their co-player might vary if they perform better or worse. We added measures of performance using overall, average, minimum, and maximum score. In addition, we added scores for each round, absolute and relative score difference between rounds as measures of the development of performance over time, score for each role, and absolute and relative score difference between roles as a measure of role-based performance.

\textbf{In-Game Behaviour:}
We used two features as a measure of the amount of in-game actions that players performed: number of horizontal and vertical pushes of game board tiles.
\begin{table}[t]
\footnotesize
\begin{tabularx}{\columnwidth}{lX}
\toprule
\textbf{Category} & \textbf{Name}\\
\midrule
chronemics & TimeSpeaking, CountSpeechSegments, CountConversationalTurns, AvgSpeechSegmentLength, AvgPauseSegmentLength, SDSpeechSegmentLength, IsDominantSpeakTime, IsDominantConvTurns, TimeSilence, FractionTimeSilence, AverageSilenceLength, FirstSilenceLength\\
\rule{0pt}{3ex} 
comm. content & CountTotalWords, CountWordsAnalytic, CountWordsClout, CountWordsAuthentic, CountWordsTone, CountWordsPronounI, CountWordsPronounWe, CountWordsPronounYou, CountWordsNumber, CountWordsAffect, CountWordsPosEmo, CountWordsNegEmo, CountWordsSocial, CountWordsAffilitation, CountWordsMotion, CountWordsSpace, CountWordsTime\\
\rule{0pt}{3ex} 
eye blink & MeanBlinkRate, StdBlinkRate\\
\rule{0pt}{3ex} 
in-game behaviour & CountVerticalPushes, CountHorizontalPushes\\
\rule{0pt}{3ex} 
fac. expr. & EngagementMean, EngagementPeaks, ContemptMean, ContemptPeaks, SurpriseMean, SurprisePeaks, AngerMean, AngerPeaks, SadnessMean, SadnessPeaks, DisgustMean, DisgustPeaks, FearMean, FearPeaks, JoyMean, JoyPeaks\\
\rule{0pt}{3ex} 
performance & ScoreRound1, ScoreRound2, ScoreCollector, ScorePusher, ScoreDiffRounds, ScoreAbsDiffRounds, ScoreDiffRole, ScoreAbsDiffRole, ScoreOverall, ScoreMean, ScoreMin, ScoreMax\\
\rule{0pt}{3ex} 
self-report & Age, GamerIdentification, GenrePuzzles, GenreCasual, SameGenderCoPlayer, Gender, GenderCoPlayer, Extraversion, Agreeableness, Conscientiousness, EmotionalStability, Openness, PropensityToTrust, BrainhexSocializer\\
\bottomrule
\end{tabularx}
\caption{Names and associated categories of all \nfeatures features.}
\label{tab:featureList}
\end{table}

\textbf{Self-Report Traits:}
We added 14 features that are based on trait-based self-report measures. We used five features representing the Big 5 personality traits, %
and a single feature for propensity to trust, as they are important for the perception of social interactions \cite{Depping:2016:TMS:2967934.2968097}. 
Gaming preference features used were self-identification as a gamer and boolean features for enjoyment of casual games and puzzle games (i.e., \labyrinth's genres) and BrainHex class of Socializer.
Finally, we added features for player age and gender, co-player gender, and gender pairing (boolean same/mixed).

\subsection{Training}

After feature generation, the dataset consisted of \nsamples samples, each representing a single participant with a vector of \nfeatures features. We trained classifiers and regression models (regressors) for the prediction of binary affiliation and continuous level of affiliation, respectively. To select suitable models, we trained and tested models using the approach outlined below. We tested Naive Bayes, Logistic Regression, Random Forests (\rf), and Support Vector Machines (\svm) as classifiers and linear regression, LASSO regression, \rf, and \svm models as regressors. Models using \rf \cite{breiman2001random, kononenko2007machine} and \svm \cite{cortes1995support} performed best, thus we chose them as models to evaluate our research questions. %

Before training, invalid data were treated as missing and estimated using multivariate feature imputation based on \emph{MICE} \cite{buuren2010mice}, with ridge regression models predicting missing data using the other features. Hyper parameters for all models were determined using repeated 10-fold cross-validated grid search (see Table~\ref{tab:paramGrid} for parameter grid). Best-performing parameters were subsequently used to train the models. 128 trees were used in the \rf models as suggested in previous work \cite{oshiro2012many} and were fully expanded. The \svm models used a radial basis function as kernel. We employed \emph{Scikit-learn} 0.20 \cite{scikit-learn} in our implementation.

Training was performed using leave-2-groups-out cross-validation (\cv). We separated 4 samples from 2 dyads from the data set as a test set and trained a model on the remaining 42 samples---repeated for all possible combinations of selecting 2 dyads as test set. We employed this approach as it covers the whole data set and provides a good trade-off of bias and variance due to split size---close to k-folds with $k=10$. We used this approach over traditional k-fold \cv as it keeps samples of dyads separate on training/test splits, and instead of leave-one-out, which has increased variance \cite{kohavi1995study}. The high computational cost was not an issue due to our comparably small number of samples. We repeated the \cv 10 times to reduce variance estimates for models, which can be a problem with small sample sizes (cf.~\cite{BELEITES200591}). With this \cv approach, we trained a completely new model for each training set and tested it on data that is unknown to the model---in contrast to approaches that refine a single model with train and test data and thus require a separate holdout set. As such, our \cv approach allows an assessment of out-of-sample prediction, i.e., how well a model using the same features could predict affiliation on similar data. Therefore, if predictions are better than random chance with our cross-validation approach, it is likely predictions are equally accurate with similar data not present in our data set.

\begin{table}[t]
\footnotesize
\begin{tabularx}{\columnwidth}{@{}lX@{}}
\toprule
\textbf{Random Forests}\\
\midrule
max features $\in \{\emph{\#features}, \emph{sqrt(\#features)}, \emph{log2(\#features)}\}$\\
min samples leafs $\in \{1, 2, 4, 8, 16\}$\\
min samples split $\in \{2, 4, 8, 16\}$\\
\midrule
\textbf{Support Vector Machines}\\
\midrule
C $\in \{0.01, 0.1, 1, 10, 100, 1000, 10000\}$\\
$\gamma$ $\in \{0.01, 0.1, 1, 10, 100, 1000, 10000\}$\\
\bottomrule
\end{tabularx}
\caption{Grid used in cross-validated grid search to determine best parameters of \rf and \svm models. \vspace{-4pt}}
\label{tab:paramGrid}
\end{table}

To gain insights into the relevance of features, we trained \rf regressors on the whole data set with recursive feature elimination using the same cross-validation approach (cf.~\cite{guyon2002gene, sklearnrfecv}). This method recursively trains models with smaller feature subsets until an optimal solution is reached. This optimal solution was reached with 9 features (see Table~\ref{tab:mostImportantFeatures}). We used a model containing these features (\bestfeatures model) in addition to the other models as a possible best model. 
In addition, we trained models with subsets of features for each feature category to test if a single category suffices, e.g., when there is no access to video data. %

\vspace{-4pt}
\section{Results}

\begin{figure}[t]
    \centering
	\includegraphics[width=.85\columnwidth]{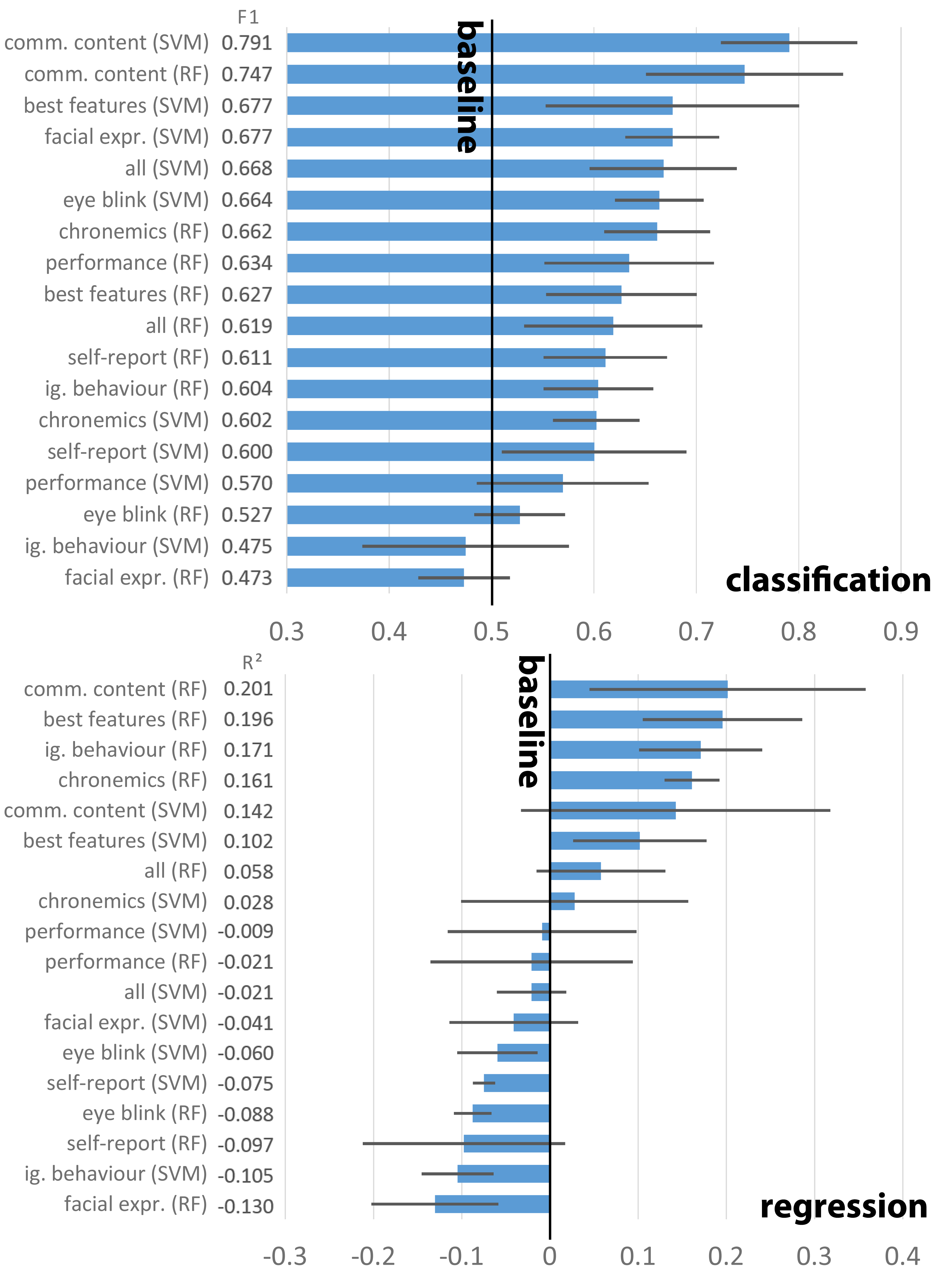}
	\caption{Average ($\pm$ \emph{SD}) \emph{F1} and $R^2$ scores for \rf and \svm classification and regression models. The baseline line depicts performance of random guess (classification) and predicting data mean (regression).}
	\label{fig:performance}
\end{figure}

\rqone and \rqtwo concern model performance. In particular, we are interested if affiliation can be predicted with a model using our features in general (\rqone) and with models using features from single categories (\rqtwo). In both cases, we compare performance to baselines. To the best of our knowledge, no previous work has tried to predict self-reported affiliation based on behavioural traits of players. %
Thus, we evaluated model performance in comparison to baselines that do not use our feature set. If a model performs better than its baseline, the combination of features has value for the prediction of affiliation. For binary classification, the baseline was random guess (\fone = .50). For the regression models, we used prediction of data set mean~\cite{dummyRegression} (\rtwo~=~0.00) as baseline as it outperformed median prediction and random guess is inappropriate for continuous data. Figure~\ref{fig:performance} shows performance measures for \rf and \svm models for the prediction of binary (\fone) and continuous (\rtwo) affiliation respectively.

To account for variance in model performance, we used statistical tests comparing performance to the baseline. 
However, frequentist t-tests and ANOVAs are not appropriate for this comparison, because the measures for a model are not independent from one another when gathered with repeated CV (cf.~\cite{bouckaert2003choosing, demvsar2006statistical, dietterich1998approximate, vanwinckelen2012estimating}). %

\subsection{Bayesian Analysis: A Primer and Description}

To avoid the potential issues of using frequentist hypothesis tests for comparing classifier performance, we followed the recent recommendation of Benavoli et al.~\cite{benavoli2017time}, who proposed using Bayesian analysis. We compared models using Bayes factors, which are a popular method of Bayesian hypothesis testing \cite{quintana2018bayesian}.
Bayes factors are a ratio of the likelihood of observing some specific data under two statistical models~\cite{wagenmaker2018part1}; in other words, they measure the ratio of the likelihood of data occurring given a null hypothesis $H_0$ and an alternative hypothesis $H_1$.
The evaluation of whether a model is better than a baseline can be formulated as a hypothesis $H_1$, how likely its accuracy measures are higher than the baseline score, which can then be tested with a Bayesian t-test. For example, data is 6.33 times more likely under $H_1$ than under $H_0$ for a hypothetical Bayes factor $BF_{10}$ equaling 6.33 \cite{wagenmaker2018part1}. %
For Bayesian analyses, we used \emph{JASP} \cite{JASP2018}, a graphical tool providing the Bayesian equivalents to one sample t-tests using an implementation of the JZS t-test as described by Rouder et al.~\cite{rouder2009bayesian}. In our analyses, we used objective, default Cauchy priors centered around 0 with a width of $1/\sqrt{2}$ and evaluated their robustness using different sensible priors widths \cite{carlsson2017bayes, quintana2018bayesian}. For all Bayes factors, we report raw $BF_{10}$. We use the interpretation of \emph{JASP} \cite{wagenmaker2018part2}, which is based on earlier interpretations  \cite{jeffreys1961theory, lee2013bayesian}, to provide context. For example, a $BF_{10}$ between 10 and 30 provides strong evidence for $H_1$ over $H_0$. In addition, we report posterior estimates for the effect sizes with median Cohen's \emph{d}.%

\subsection{RQ1: Recognition of Affiliation}
In \rqone, we ask: \emph{Is it possible to predict a player's interpersonal affiliation toward a co-player in an online multiplayer game setting from unobtrusively gathered behavioural data?}  %
The results suggest performance does vary largely with respect to the selected features and model. For \rqone, we are interested in models with features from multiple categories, i.e., the \emph{all} and \emph{best features} models. For these models, the \svm models performed better than \rf models for classification, while performance was better for \rf regressors than for \svm regressors. Unsurprisingly, the \bestfeatures models were better than models using all features as they disregard potentially uninformative features. Disregarding single category models temporarily, the best general models were \svm \bestfeatures for classification and \rf \bestfeatures for regression. We compared these to the baseline for \rqone. Table~\ref{tab:ttestresults} shows results of Bayesian one-sample t-tests comparing all models to their respective baselines.

A Bayesian one-sample t-test tested $H_1$, that \fone scores of the \svm \bestfeatures classifier were higher than the random guess baseline (\fone = 0.5). The results suggest \emph{very strong} ($BF_{10} > 30$) evidence of the data for $H_1$ over $H_0$ (\bfval{63.23}, median \cohensd{1.22}). In fact, the other models with \bestfeatures and even \emph{all} features might be more suitable due to lower variance. This shows that models using behavioural traces are better than the random guess baseline with 67.7\% accuracy ($F_1$) for the \bestfeatures model, suggesting that predicting binary affiliation is possible with these features.

Similarly, a Bayesian one-sample t-test tested $H_1$, that \rtwo scores of the \rf \bestfeatures regressors were higher than baseline regression performance (\rtwo = 0.0). The results provided \emph{extreme} ($BF_{10} > 100$) evidence of the data for $H_1$, i.e., that \rtwo scores were higher than baseline (\rtwo = 0.0). For the regression models, this performance was highest with 19.6\% explained variance (not considering single category models). This shows that predicting continuous affiliation is better with our features than just predicting mean affiliation score, which suggests that predicting continuous affiliation is possible as well.

In summary, the data suggest that our models can predict binary and continuous affiliation better than chance, indicating that an evaluation of social interaction quality using behavioral traces is possible. This means that a game can generate features for a gaming session and use these to predict how players experienced the interaction with other players. This works without any additional input from humans, allowing extensive insights into social player experience, while also allowing researchers to use this information in automated systems, such as for improved matchmaking.

\begin{table}[ht]
\scriptsize
\begin{tabularx}{\columnwidth}{lrrrr}
\toprule
                       & \multicolumn{2}{c}{\textbf{Classification}}             & \multicolumn{2}{c}{\textbf{Regression}}\\ \midrule
                       & \multicolumn{1}{r}{$BF_{10}$}                  & \multicolumn{1}{r}{median \emph{d}} & \multicolumn{1}{r}{$BF_{10}$}           & \multicolumn{1}{r}{median \emph{d}}\\ \midrule
\textit{all (RF)} & 51.37 & 1.14 & 5.02 & 0.67\\
\textit{all (SVM)} & 1377.92 & 2.16 & 0.14 & 0.10\\
\textit{best features (RF)} & 193.90 & 1.51 & 815.38 & 1.94\\
\textit{best features (SVM)} & 63.23 & 1.22 & 49.17 & 1.18\\
\textit{chronemics (RF)} & 11530.00 & 3.02 & 604314.44 & 5.26\\
\textit{chronemics (SVM)} & 1936.18 & 2.25 & 0.56 & 0.27\\
\textit{comm. content (RF)} & 2546.07 & 2.37 & 36.61 & 1.09\\
\textit{comm. content (SVM)} & 130617.05 & 4.29 & 5.22 & 0.68\\
\textit{eye blink (RF)} & 2.47 & 0.54 & 0.01 & 0.08\\
\textit{eye blink (SVM)} & 49321.29 & 3.70 & 0.10 & 0.12\\
\textit{facial expr. (RF)} & 0.13 & 0.09 & 0.03 & 0.09\\
\textit{facial expr. (SVM)} & 55007.78 & 3.80 & 0.14 & 0.11\\
\textit{in-game behaviour (RF)} & 438.08 & 1.80 & 1958.32 & 2.19\\
\textit{in-game behaviour (SVM)} & 0.19 & 0.13 & 0.02 & 0.11\\
\textit{performance (RF)} & 135.19 & 1.44 & 0.22 & 0.15\\
\textit{performance (SVM)} & 5.78 & 0.70 & 0.26 & 0.17\\
\textit{self-report (RF)} & 304.15 & 1.66 & 0.12 & 0.07\\
\textit{self-report (SVM)} & 18.51 & 0.96 & 0.01 & 0.14\\
\midrule
\textit{$H_1$}         & \multicolumn{2}{c}{\textit{$F_1$ higher than 0.5}} & \multicolumn{2}{c}{$R^2$ \textit{higher than 0.0}} \\ 
\bottomrule
\end{tabularx}
\caption{Bayes factors and posterior effects for one sample t-tests comparing models to their respective baselines. $BF_{10}$ indicates support for $H_1$ over $H_0$, i.e., for $H_1$  values over 1 and for $H_0$ for values lower than 1. %
}
\label{tab:ttestresults}
\end{table}
 \begin{table}[th]
 \footnotesize
 \begin{tabularx}{\columnwidth}{XXr}
 \toprule
 \textbf{Feature} & \textbf{Category} & {$r_\tau$}\\
 \midrule
 AvgPauseTime      & \chronemics & $-.259$\\
 CountConversationalTurns & \chronemics & $.328$\\
 CountWordsAnalytic  & \commcontent & $-.288$\\
 CountWordsNumber    & \commcontent & $-.287$\\
 CountWordsPronounI  & \commcontent & $.276$\\
 CountWordsTime      & \commcontent & $.333$\\
 CountHorizontalPushes      & \events & $.066$\\
 PropensityToTrust   & \selfreport & $.169$\\
 Conscientiousness   & \selfreport & $.071$\\
 \bottomrule
 \end{tabularx}
 \caption{The most important features as determined by cross-validated recursive feature elimination and Random Forest regressors predicting continuous affiliation. Kendall tau-B correlations indicate potential direction of relation between variables.}%
 \label{tab:mostImportantFeatures}
 \end{table}

 \vspace{-4pt}
\subsection{RQ2: Using Single Category Models}

\rqtwo asks: \emph{How do models using only features from a single category perform?} This is interesting as a single-category model would allow the evaluation of social interactions even if researchers have access only to specific data streams, such as players' voice chat or even only in-game data. This type of model could be desirable because not all data sources might be available for each game context or might not be accessible at all, due to restrictions related to privacy or ethics (see Discussion). We tested models using only features from each category to investigate the performance of single-category models. Table~\ref{tab:ttestresults} shows the results of the Bayesian t-tests comparing performance to the baselines. 

Regarding classification, \rf models showed promise for models using in-game data (\emph{in-game behaviour} \& \emph{performance}), whereas \svm classifiers outperformed \rf classifiers for the features gathered from video data (\emph{eye blink} \& \emph{facial expression}). Overall, the results suggest that for each category, there is a model that has acceptable accuracy, suggesting that single-category models might be useful to varying degrees. The best results were achieved by the models based on verbal communication (\commcontent \& \chronemics), where the t-tests strongly suggested better than baseline performance. Performance measures for the \svm \commcontent model suggested the highest likelihood of being better than the baseline, indicating best performance for classification ($F_1 = 0.791$) and outperforming the \bestfeatures model.

Results were more varied for regression models. Bayesian one-sample t-tests suggested evidence that 7 models performed better than the baseline. The %
tests suggested evidence for $H_0$, i.e., that performance was not better than baseline for 11 models, including for all video-based feature sets (\emph{eye blink} \& \emph{facial expression}), \selfreport features, most in-game data feature sets, and even the \svm \emph{all} feature set. The worse performance in comparison to the classification task can be explained by the higher difficulty of predicting continuous affiliation and the better baseline method (predicting mean vs.\ random guess). \rtwo scores below zero are caused by a model that does not predict well on the test set. This suggests that in these cases, the models did overfit on the training data and a general relationship is unlikely, leading to unsuitable predictions on new data. %
On the other hand, models with \bestfeatures and \commcontent as well as the \rf regressors using \events and \chronemics performed better than the baseline. Results suggest the best performance for \rf regressor using \commcontent ($R^2=0.201$) and \chronemics ($R^2=0.161$ and lower variance). %

\subsection{RQ3: Feature Importance}
\rqthree asks: \emph{Which behavioural traces are important features for predicting affiliation during play?}
To gather first insights on the relationship between player behavior and affiliation, we evaluated feature importances. We consider features that are important for prediction as potential indicators of affiliation. %
Table~\ref{tab:mostImportantFeatures} shows the 9 features of the optimal feature set as determined by the cross-validated recursive feature elimination with \rf regressors. 

Higher affiliation coincided with more communication in the form of lower overall conversational pause time and a higher number of conversational turns. Four \commcontent features were also in the set of important features. Higher affiliation was reflected in fewer words related to analytic thinking and numbers as well as greater usage of the pronoun \emph{``I''}---including variations like \emph{``me''}---and words related to time. %
Further analysis is needed to characterize the relationship between communication content and affiliation; however, we speculate that use of analytic language and numbers, such as to discuss scores, reflects players manifesting their motivation to perform well as explicit discussion, rather than relying on their partner, engendering less affiliation.
This lies in contrast, however, to the positive relationship between words related to time and affiliation, considering that time limits were an important part of gameplay. The positive correlation of affiliation and \emph{``I''}-related words might be related to players communicating their actions, leading to greater feelings of being a team. It could even suggest that players may have revealed personal information about themselves to the other player, which is in line with research on privacy suggesting a link between trust and self-disclosure \cite{joinson2010privacy}. Interestingly, the number of horizontal pushes was included in the optimal feature set, whereas scores were only slightly correlated. This suggests that considering features of the \emph{process} of playing can be valuable for prediction when used in combination with other features. %
Finally, in line with previous work on personality traits and trust \cite{Depping:2016:TMS:2967934.2968097} and trustworthiness \cite{evans2008survey}, higher affiliation coincided with higher propensity to trust and conscientiousness. Keep in mind that the reported correlation scores are not corrected for dyads and are therefore overestimated. Due to the exploratory nature of this $RQ$, we did not want to conduct analyses controlling for the relationship amongst features, as this would lead to unreliable estimates of effects and significance that could be misinterpreted. %
We report these feature importances to give an overview of the direction of a relationship, informing future work with controlled experiments, while our results do not reflect a deeper understanding of the connection between features and affiliation. %

\section{Discussion}
We discuss findings, generalizability, and application.

\subsection{Summary of Findings}
We summarize our findings as follows: (1) Affiliation can be predicted from player behaviour. Our results show that the prediction of both binary and continuous affiliation is possible with up to 79.1\% accuracy and 20.1\% explained variance. (2) The best models strongly outperformed the baseline models, suggesting that reliable recognition of social interaction quality based on behavioral traces is possible and feasible. %
(3) Binary affiliation can be predicted with accuracy better than chance from various sets of features (2 models > 70\% accuracy, 14 better than baseline, 2 not useful). (4) Predicting continuous affiliation is possible but more challenging (4 models > 15\% explained variance, 3 better than baseline, 11 not useful). (5) Models using only \commcontent or \chronemics performed best for both classification and regression indicating value of features based on verbal communication. %

\subsection{Generalizability of Findings}
Our approach applies machine learning methods to gameplay data, i.e., human behaviour and self-reported appraisal of interpersonal interaction. Due to the deviation from experimental studies and their analysis, we provide context on findings and generalizability. %

First, with respect to \rqone, we demonstrated validity by showing that binary and continuous predictions are possible with up to 79.1\% accuracy for and 20.1\% explained variance on data unknown to the models. 
Due to potential bias in selection, we did not use a dedicated holdout set. Thus, our assessment of generalizability is limited to the cross-validated performance, which is an estimate of out-of-sample performance that could be expected for players in similar scenarios.
While we cannot provide a conclusive assessment of the generalizability, this paper suggests validity of our proposed novel approach to unobtrusively assess social interaction quality in games. 
With our cross-validation, we found that some models likely were overfit, as is common with a high number of features compared to the number of samples. For these, we suspect that they might not generalize well beyond our sample. Based on cross-validation, we suspect they perform well for similarly behaving players, but require further studies to confirm generalizability to other players. The analysis of models with fewer features (e.g., \emph{chronemics}), where overfitting is less likely, reinforces the potential generalized performance of this approach. %

Second, our intention was not to study whether and how a specific behavioural trait is related to affiliation. As such, the analysis of feature importances does not provide generalizable insights into the relationship between behaviour and affiliation. %
Based on the analysis of the feature importances, we provide a set of features that in combination is important for predicting affiliation. Correlation measures give potential insights into the relationship of the variables, but with our approach we cannot meaningfully control for interaction effects or correlations amongst these variables without overestimating effects. While we cannot draw conclusions on the general relationship between our variables, our results can be used to inform hypotheses in future controlled experiments that allow for causal inference. %

Third, some readers might wonder if better-than-chance prediction rates are good enough for real-world application. The decades of research on emotion recognition have shown that assessing complex psychological states is challenging and performance should be considered as context-dependently ``good enough''. Our paper demonstrates that predicting affiliation based on human behaviour is possible with acceptable performance suggesting validity of the approach. Real value then depends on the use case. Predictions are likely better for players known to the models and therefore are useful when applied in published games where state is predicted repeatedly, e.g., for each match. Further, models can be valuable without high accuracy but specifically trained to detect particularly bad or good experiences (i.e., low/high affiliation), because such states are more relevant for assessing and evaluating game features, e.g., if a new chat feature leads to many negatively perceived interactions.

\subsection{Detecting Social Interaction Quality in Games}

People may experience the same social interaction differently depending on their context, personality, or previous experience. Whereas approaches such as the detection of toxic behaviour \cite{martens2015toxicity, THOMPSON2017149} try to assess if a message is toxic or offensive, they mostly assume that there is a generally agreed-upon definition of what they aim to predict, i.e., a negative interaction. We argue there is no general truth of the social interaction---what one person (or algorithm) considers harmless might deeply offend another player, while a third may think it is a hilarious and integral part of the in-game interaction between friends. Further, one player may think that a comment is funny one day and hurtful the next, depending on their mood and the circumstances. An assessment of social interactions must be grounded in the appraisal of the experience of this interaction. More generally, the way in which people interact with each other is highly complex, and to capture the full degree of how each player experiences an interaction, a system needs to examine a variety of very subtle cues. For example, a player blushing after another player helped them out might be a good indicator for a positive social interaction that can be interpreted easily by a human but is difficult to assess for an algorithm. 
We propose that our work provides a first step toward the goal of evaluating the experience of social interaction in games by showing that it is possible to predict self-reported affiliation. 
While many essential questions remain to be answered, this paper establishes the potential of this approach.

\subsubsection{Application of Affiliation Recognition}
The assessment of social experiences is useful for game developers to evaluate multiplayer games. The design and development of multiplayer games is very expensive and thus carries substantial financial risk; %
our approach can help inform low-fidelity and unobtrusive measurement tools that would be valuable for games-as-a-service to assess the quality of social experiences. While we used it for a single assessment, this approach can be used to provide a continuous measurement that allows for an evaluation of the progress of social interaction quality over time and also attribute the experience of a social interaction to specific micro events by creating windows of gameplay phases. This can help game developers assess an essential aspect of multiplayer experiences for which there are surprisingly few automated, unobtrusive, and continuous measurement tools. 

In our analysis, we used an offline feature generation pipeline, but all steps could be applied online in real-time. In a real application, the prediction models can be pre-trained and improved over time. The same feature generation pipeline generates feature vectors that the model uses as input. Predictions are not costly. Implementing a pipeline as described allows an automated, continuous, real-time evaluation of player states. %

In this paper, we propose the behaviour-based assessment of social interaction quality, but not how to use this knowledge in a game to improve the interaction between players. Improving problematic interpersonal interactions is a difficult problem and an easy solution might not exist. %
Developers of multiplayer games (similar to other platforms on which humans interact) need to address hostile, toxic, and otherwise negative interpersonal interactions, which is a challenging and perpetual task. Our work demonstrates that affiliation effectively can be predicted based on behaviour, and further provides guidance on feature subsets that are particularly salient. We hope that game developers can use our findings and that our work helps contribute to a shared effort of industry practitioners and academic researchers to create healthier, more positive environments for players, in which the risk of negative and toxic interactions is minimized.

In addition to measuring the quality of social interactions to inform design and development of games and game communities, our findings have interesting applications in adaptive gaming. 
In many multiplayer games, players are assigned to teams with strangers by matchmaking systems. Games that are able to detect the quality of social interactions as experienced by players can directly leverage this information without requiring additional explicit input from players. A system that can automatically detect the state of a player can react accordingly, or better yet, can preventatively act when it predicts that a player could experience a state before it even happens. Adaptation and preventative action are major benefits of using an automated assessment of how players experience a social interaction. A system could prevent toxicity in a game community by giving appropriate feedback to players whose behavior is experienced as hostile or by protecting victims of negative behavior, e.g., by censoring hurtful messages with an approach that is aware of how players perceive these messages. As such, a computational model in combination with an adaptive system that knows how to handle varying quality of social interactions can not only assess experiences but provide user-specific solutions like directed feedback or matchmaking that finds suitable partners, while acknowledging that people differ and sometimes interpret and experience the same social interaction from very different perspectives.

\subsubsection{Group Dynamics}
Our overall goal is to recognize the quality of social interactions in multiplayer games. In this paper, we avoided group dynamics and operationalized social interaction quality as affiliation between dyads in a cooperative interdependent game. In commercial multiplayer games, there exist a variety of other contexts and considerations. While we suspect that it is still possible to assess an individual's---as well as the group's---overall experience of the social interaction in settings with multiple co-players, this requires future research. In particular, it seems challenging to account for group dynamics and attribute a player's experience to specific co-players. Similarly, we suspect that our findings are in part limited to the cooperative setting that we employed and a generalization to the quality of social interactions with opponents in competitive settings requires further research. We think that a recognition of the quality of social interactions in groups settings can be challenging, but could benefit games that struggle with negative social experiences such as toxicity.

In our approach, we predict unidirectional affiliation and consider behavioural traces of a player as indicative of their affiliation for an initial investigation of feasibility, while mostly disregarding the other player. We suspect that performance can be improved by leveraging the fact that a player's appraisal of the quality of a social interaction is probably dependent on the behaviour from all involved players. A player's affiliation toward a teammate is affected by the things they say, e.g., when they use supportive or hurtful language. We suspect that models that use features from all involved players in an interaction are better suited to evaluate the quality of an interaction as suggested by the strong performance with communication features. As such, we think that models with information of all players predicting affiliation for all of them is a promising direction for future work.

\subsubsection{Privacy and Ethical Considerations}
Our method relies on collecting behavioural data, which can affect players' privacy. First, it is important that such methods are only used sparingly and with informed consent of the users. Different types of data affect privacy differently. Features based on audio and video streams rely on the analysis of player behaviour in the physical world, which encompasses unrelated activities during gameplay (e.g., eating), the surrounding environment (e.g., letters on a desk), or other people (e.g., family members talking in the background). While it is important that all data is treated ethically, there is a higher danger of mismanagement with such data than relying on less critical data such as those based on in-game behaviours and performance.
As there are trade-offs between functionality and privacy~\cite{domingo2009functionality}, affecting players' privacy is not always worth it. While this privacy intrusion might be worth to prevent hurtful messages, collecting video data for better matchmaking is potentially unnecessarily invasive.
As such, it is important that developers consider these trade-offs and, if possible, use less invasive features such as in-game data. 
Generally, raw data should not be stored on centralized servers. Instead, feature generation and models can be implemented on the client-side in anonymized form, which poses fewer problems for privacy. However, a game that predicts affiliation from player behaviour needs to be certain to provide a benefit for players and only use such data with the explicit informed consent of players.

\vspace{-6pt}
\section{Conclusion}
This paper provides evidence that behavioural traces can be used to reliably predict players' affiliation toward a co-player in a dyadic cooperative online game. Our results suggest this is possible to varying degrees with many different types of features, i.e., with models using only chronemics, communication content, in-game events, or in-game performance features. %
This work can assist game developers by building toward a powerful, automated, and continuous method of evaluating the quality of social interactions as they are experienced by players. %

\vspace{-6pt}
\section{Acknowledgments}
We thank NSERC and SWaGUR for funding, members of our labs for feedback, and our participants.

\balance{}

\bibliographystyle{SIGCHI-Reference-Format}
\bibliography{bibliography}

\end{document}